\documentclass[a4paper,onecolumn]{article}
\usepackage[pdftex]{graphicx}
\pdfoutput=1
\usepackage{latexsym}
\usepackage[fleqn]{amsmath}
\usepackage[psamsfonts]{amssymb}
\usepackage{amsthm}

\def\x{\boldsymbol{x}}
\def\y{\boldsymbol{y}}

\def\0{\boldsymbol{0}}

\def\cB{\mathcal B}

\def\cI{\mathcal I}

\def\cM{\mathcal M}

\def\cP{\mathcal P}
\def\cR{\mathcal R}
\def\cS{\mathcal S}
\def\cT{\mathcal T}

\def\cV{\mathcal V}

\def\cX{\mathcal X}
\def\cY{\mathcal Y}

\def\ov{\overline}
\def\ul{\underline}
\def\wh{\widehat}

\def\eqdef{{\displaystyle\mathop{=}^{\mbox{\rm def.}}}}

\newcommand\dotdot[1]{\mbox{$\ddot{\mbox{#1}}$}}
\newcommand\dash[1]{\mbox{$\acute{\mbox{#1}}$}}%

\theoremstyle{plain}
\newtheorem{teiri}{Theorem}
\newtheorem{kei}{Corollary}
\newtheorem{hodai}{Lemma}
\theoremstyle{definition}
\newtheorem{teigi}{Definition}
\theoremstyle{plain}

\title{%
  Universal coding for correlated sources with complementary delivery
  \thanks{%
    Some of the material in this manuscript has been already published in IEICE
    Transactions on Fundamentals, Vol.E90-A, No.9, pp.1840-1847, September 2007.
    Manuscript received Derember 19, 2006.
    Manuscript revised March 29, 2007.
    Final manuscriot received April 20, 2007.
    Several additional results are also included.
  }
  \thanks{%
    Parts of the material in this paper were presented at the International Symposium on
    Information Theory, Nice, France June 2007.
  }%
}
\author{%
  Akisato Kimura
  \thanks{%
    NTT Communication Science Laboratories, NTT Corporation,
    3-1 Morinosato Wakamiya, Atsugi-shi, Kanagawa, 243-0198 Japan.
    E-mail: research $<$at$>$ akisato org
    URL: http://www.brl.ntt.co.jp/people/akisato/
  }%
  \and Tomohiko Uyematsu
  \thanks{%
    Department of Communications and Integrated Systems, Tokyo Institute of Technology,
    2-12-1 Ookayama, Meguro-ku, Tokyo, 152-8550 Japan.
    E-mail: uyematsu $<$at$>$ ieee org
  }%
  \and Shigeaki Kuzuoka
  \thanks{%
    Department of Computer and Communication Sciences, Wakayama University.
    930 Sakaedani, Wakayama-shi, Wakayama, 640-8510 Japan.
    E-mail: kuzuoka $<$at$>$ sys wakayama-u ac jp
  }%
}
\begin{document}
\allowdisplaybreaks  % allow to repage in the mathematical environments

\maketitle

\begin{abstract}
This paper deals with a universal coding problem for a certain kind of multiterminal
source coding system that we call the complementary delivery coding system. In this
system, messages from two correlated sources are jointly encoded, and each decoder has
access to one of the two messages to enable it to reproduce the other message. Both
fixed-to-fixed length and fixed-to-variable length lossless coding schemes are
considered. Explicit constructions of universal codes and bounds of the error
probabilities are clarified via type-theoretical and graph-theoretical analyses.
\end{abstract}

%%%%%%%%
\section{Introduction}

The coding problem for correlated information sources was first described and
investigated
by Slepian and Wolf \cite{SlepianWolf}, and later, various coding problems derived from
that work were considered (e.g. Wyner \cite{SideInformationCoding:Wyner}, K\dotdot{o}rner
and Marton \cite{KornerMarton}, Sgarro \cite{SgarroCoding}). Meanwhile, the problem of
universal coding for these systems was first investigated by Csisz\dash{a}r and
K\dotdot{o}rner \cite{UniversalSlepianWolf:Csiszar}. Universal coding problems are not
only interesting in their own right but also very important in terms of practical
applications. Subsequent work has mainly focused on the Slepian-Wolf coding system
\cite{UniversalSlepianWolf:Csiszar2,UniversalSlepianWolf:Oohama,%
UniversalSlepianWolf:Uyematsu} since it appears to be difficult to construct universal
codes for most of the other coding systems. For example, Muramatsu \cite{PhD:muramatsu}
showed that we cannot construct asymptotically optimal fixed-to-variable length (FV)
universal code for Wyner-Ziv coding systems \cite{WynerZiv} in terms of the coding rate.

\begin{figure}[t]
  \begin{center}
    \includegraphics[width=0.975\hsize]{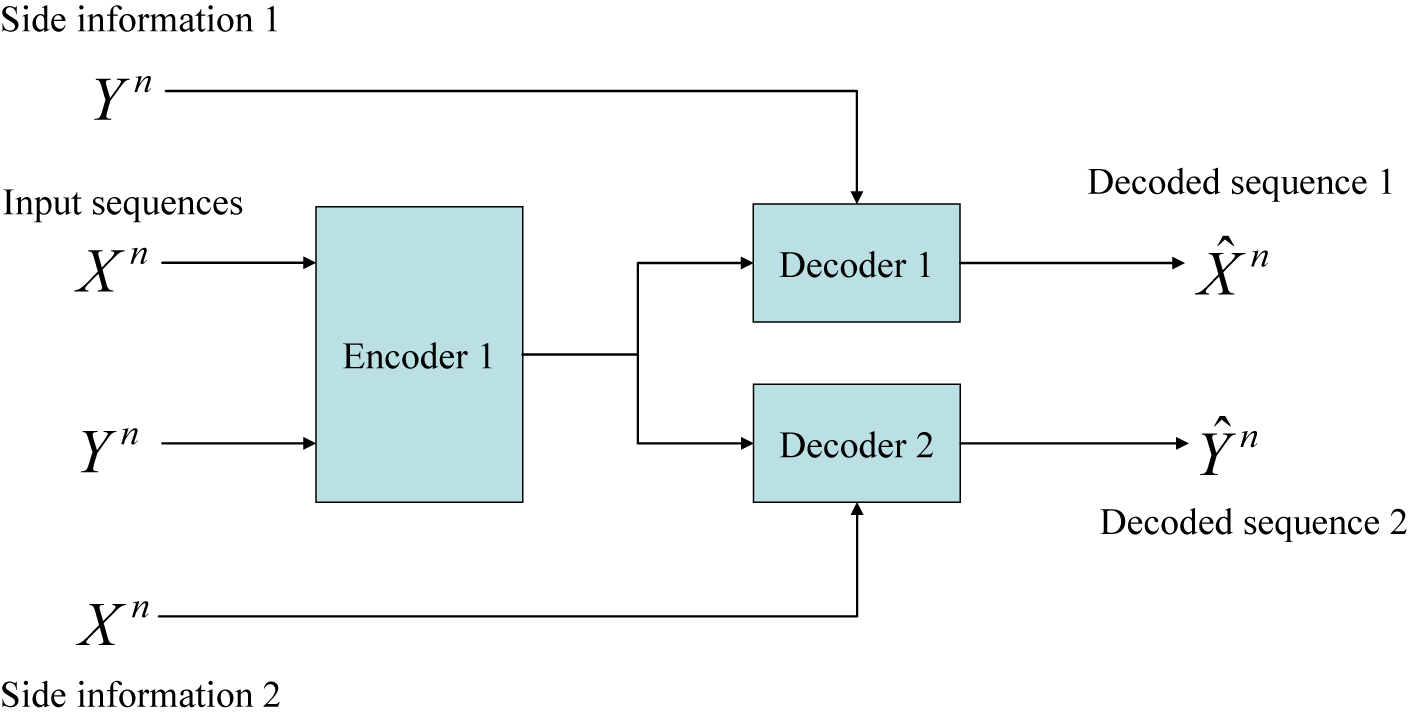}
    \caption{Complementary delivery coding system}
    \label{fig:complementary}
  \end{center}
\end{figure}

This paper deals with a universal coding problem for a certain kind of multiterminal
source coding system that we call a complementary delivery coding system
\cite{ComplementaryISITA}. Figure \ref{fig:complementary} shows a
block diagram of the complementary delivery coding system. The encoder observes messages
emitted from two correlated sources, and delivers these messages to other locations (i.e.
decoders). Each decoder has access to one of two messages, and therefore wants to
reproduce the other message.
Although the previous articles \cite{ComplementaryISITA} considered
lossy configurations, this paper considers a lossless configuration. We show an explicit
construction of fixed-to-fixed length universal codes. We also clarify the upper and
lower bounds of the error probabilities via type-theoretical and graph-theoretical
analyses. Fixed-to-variable universal codes can also be constructed in a similar manner.

This paper is organized as follows: Notations and definitions are provided in Section
\ref{sec:pre}. Previous results for the complementary delivery coding system are shown in
Section \ref{sec:previous}. A proposed coding scheme is described in Section
\ref{sec:construct}. Several coding theorems are clarified in Section \ref{sec:theorem}.
Variable-length coding is discussed in Section \ref{sec:variable}. Finally,
concluding remarks are given in Section \ref{sec:conclude}.

%%%%%%%%
\section{Preliminaries}
\label{sec:pre}

%%%%
\subsection{Basic definitions}
\label{sec:pre:defs}

Let $\cX$ be a finite set, $\cB$ be a binary set, and $\cB^*$ be a set of all finite
sequences in the alphabet $\cB$. Let $|\cX|$ be the cardinality of $\cX$ and
$\cI_M=\{1,2,\cdots,M\}$. A member of $\cX^n$ is written as $x^n=(x_1,x_2,\cdots,x_n)$,
and substrings of $x^n$ are written as $x_i^j=(x_i,x_{i+1},\cdots,x_j)$ for $i\le j$.
When the dimension is clear from the context, vectors will be denoted by boldface
letters, i.e., $\x\in\cX^n$. $\cM(\cX)$ denotes the set of all probability distributions
on $\cX$. Also, $\cM(\cX|P_Y)$ denotes the set of all probability distributions on $\cX$
given a distribution $P_Y\in\cM(\cY)$, namely each member $P_{X|Y}$ of $\cM(\cX|P_Y)$ is
characterized by $P_{XY}\in\cM(\cX\times\cY)$ as $P_{XY}=P_{X|Y}P_Y$. A discrete
memoryless source $(\cX,P_X)$ is an infinite sequence of independent copies of a random
variable $X$ taking values in $\cX$ with a generic distribution $P_X\in\cM(\cX)$. We will
denote a source $(\cX,P_{X})$ by referring to its generic distribution $P_X$ or random
variable $X$. For a correlated source $(X,Y)$, $H(X|Y)$ denotes the conditional entropy
of $X$ given $Y$. For a generic distribution $P_Y\in\cM(\cY)$ and a conditional
distribution $P_{X|Y}\in\cM(\cX|P_Y)$, $H(P_{X|Y}|P_Y)$ also denotes the
conditional
entropy of $X$ given $Y$. $D(P\| Q)$ denotes the Kullback-Leibler divergence between two
distributions $P$ and $Q$. In the following, all bases of exponentials and logarithms are
set at 2.

%%%%
\subsection{Types of sequences}
\label{sec:pre:type}

Let us define the {\it type} of a sequence $\x\in\cX^n$ as the empirical distribution
$Q_{\x}\in\cM(\cX)$ of the sequence $\x$, i.e.
\begin{eqnarray*}
  Q_{\x}(a) &=& \frac 1n N(a|\x)\quad\forall a\in\cX,
\end{eqnarray*}
where $N(a|\x)$ represents the number of occurences of the letter $a$ in the sequence
$\x$. Similarly, the joint type $Q_{\x,\y}\in\cM(\cX\times\cY)$ is defined by
\begin{eqnarray*}
  Q_{\x,\y}(a,b) &=& \frac 1n N(a,b|\x,\y)\quad\forall(a,b)\in\cX\times\cY.
\end{eqnarray*}
Let $\cP_n(\cX)$ be the set of types of sequences in $\cX^n$. Similarly, for every type
$Q\in\cP_n(\cX)$, let $\cV_n(\cY|Q)$ be the set of all stochastic matrices $V:\cX\to\cY$
such that for some pairs $(\x,\y)\in\cX^n\times\cY^n$ of sequences we have
$Q_{\x,\y}(\x,\y)=Q(\x)V(\y|\x)$. For every type $Q\in\cP_n(\cX)$ we denote
\begin{eqnarray*}
  T_Q^n &{\displaystyle\mathop{=}^{\mbox{\rm def.}}}& \{\x\in\cX^n: Q_{\x}=Q\}.
\end{eqnarray*}
Similarly, for every $\x\in T_Q^n$ and $V\in\cV_n(\cY|Q)$ we define
\begin{eqnarray*}
  \lefteqn{T_V^n(\x) {\displaystyle\mathop{=}^{\mbox{\rm def.}}} \{\y\in\cY^n: }\\
  & & Q(x)V(y|x)=Q_{\x,\y}(x,y),~\forall(x,y)\in\cX\times\cY\}.
\end{eqnarray*}
Hereafter, we call $T_V^n(\x)$ a {\it V-shell}.

Here, let us introduce several important properties of types.

\begin{hodai} \label{lemma:typecount}
  {\rm (Type counting lemma \cite[Lemma~2.2]{CsiszarKorner})}\\
  The number of different types of sequences in $\cX^n$ is less than $(n+1)^{|\cX|}$,
  namely
  \begin{eqnarray*}
    |\cP_n(\cX)| &\le& (n+1)^{|\cX|}.
  \end{eqnarray*}
\end{hodai}

\begin{hodai} \label{lemma:sizeshell}
  {\rm (Sizes of V-shells \cite[Lemma~2.5]{CsiszarKorner})}\\
  For every sequence $\x\in\cX^n$ and every stochastic matrix $V: \cX\to\cY$ such that
  the corresponding V-shell $T_V(\x)$ is not empty, we have
  \begin{eqnarray*}
    |T_V(\x)| &\ge& (n+1)^{-|\cX||\cY|}\exp\{nH(V|P_X)\},\\
    |T_V(\x)| &\le& \exp\{nH(V|P_X)\}.
  \end{eqnarray*}
\end{hodai}

\begin{hodai} \label{lemma:prob}
  {\rm (Probabilities of types \cite[Lemma~2.6]{CsiszarKorner})}\\
  For every type $Q$ of sequences in $\cX^n$ and every distribution $P_X$ on $\cX$,
  we have
  \begin{eqnarray*}
    P_X(\x)  & = & \exp\{-n(D(Q\| P_X)+H(Q))\}\quad\forall\x\in T_Q,\\
    P_X(T_Q) &\ge& (n+1)^{-|\cX|}\exp\{-nD(Q\| P_X)\},\\
    P_X(T_Q) &\le& \exp\{-nD(Q\| P_X)\}.
  \end{eqnarray*}
\end{hodai}

%%%%
\section{Previous results}
\label{sec:previous}

This section formulates the coding problem investigated in this paper, and
shows a fundamental bound of the coding rate. We again note that previous work
\cite{ComplementaryISITA} considered lossy coding, whereas this paper
considers lossless coding.

First, we formulate the coding problem of the complementary delivery coding system.

\begin{teigi}  \label{def:code}
  {\rm (Fixed-to-fixed complementary delivery (FF-CD) code)
   \cite{ComplementaryISITA}}\\
  A set $(\varphi^n,\wh{\varphi}_{(1)}^n,\wh{\varphi}_{(2)}^n)$ of an encoder and two
  decoders is an FF-CD code with parameters $(n$, $M_n$, $e_n^{(X)}$, $e_n^{(Y)})$
  for the source $(X,Y)$ if and only if
  \begin{eqnarray*}
    && \varphi^n            : \cX^n\times\cY^n\rightarrow\cI_{M_n}\\
    && \wh{\varphi}_{(1)}^n : \cI_{M_n}\times\cY^n\rightarrow\cX^n,~
       \wh{\varphi}_{(2)}^n : \cI_{M_n}\times\cX^n\rightarrow\cY^n,\\
    && e_n^{(X)} = \Pr\left\{X^n\neq\wh{X}^n\right\},~
       e_n^{(Y)} = \Pr\left\{Y^n\neq\wh{Y}^n\right\},
  \end{eqnarray*}
  where
  \begin{eqnarray*}
    \wh{X}^n &{\displaystyle\mathop{=}^{\mbox{\rm def.}}}&
      \wh{\varphi}_{(1)}^n(\varphi^n(X^n,Y^n),Y^n),\\
    \wh{Y}^n  &{\displaystyle\mathop{=}^{\mbox{\rm def.}}}&
      \wh{\varphi}_{(2)}^n(\varphi^n(X^n,Y^n),X^n).
  \end{eqnarray*}
\end{teigi}

\begin{teigi}  \label{def:rate}
  {\rm(FF-CD-achievable rate)}\\
  $R$ is an FF-CD-achievable rate of the source $(X,Y)$ if and only if there exists a
  sequence $\{(\varphi^n,\wh{\varphi}_{(1)}^n,\wh{\varphi}_{(2)}^n)\}_{n=1}^{\infty}$
  of FF-CD codes with parameters $\{(n,M_n,e_n^{(X)},e_n^{(Y)})\}_{n=1}^{\infty}$
  for the source $(X,Y)$ such that
  \begin{eqnarray*}
    && \limsup_{n\to\infty}\frac 1n\log M_n \le R,\\
    && \lim_{n\to\infty}e_n^{(X)} = \lim_{n\to\infty}e_n^{(Y)} = 0.
  \end{eqnarray*}
\end{teigi}

\begin{teigi}  \label{def:min_rate}
  {\rm(Inf FF-CD-achievable rate)}
  \begin{eqnarray*}
    \lefteqn{R_f(X,Y) = \inf\{R:}\\
    & & R \mbox{ is an FF-CD-achievable rate of }(X,Y)\}.
  \end{eqnarray*}
\end{teigi}

Willems, Wolf and Wyner \cite{BroadcastSatelliteCodingOrig,BroadcastSatelliteCoding}
investigated a coding problem where several users are physically separated but
communicate with each other via a satellite, and determined the minimum coding rate for
the three users when transmitting to and from the satellite. The complementary delivery
coding system is a special case of the system described by Willems et al., which
considers the case of two users. Therefore, we can immediately obtain the closed form of
$R_f(X,Y)$ from the result obtained by Willems et al.
\begin{teiri}  \label{theorem:complementary}
  {\rm (Coding theorem of FF-CD codes)}
  \begin{eqnarray*}
    R_f(X,Y) &=& \max\{H(X|Y),H(Y|X)\}\\
             &=& \max\{H(P_{X|Y}|P_Y),H(P_{Y|X}|P_X)\}
  \end{eqnarray*}
\end{teiri}

%%%%%%%%
\section{Code construction}
\label{sec:construct}

\begin{figure}[t]
  \begin{center}
    \includegraphics[width=0.325\hsize]{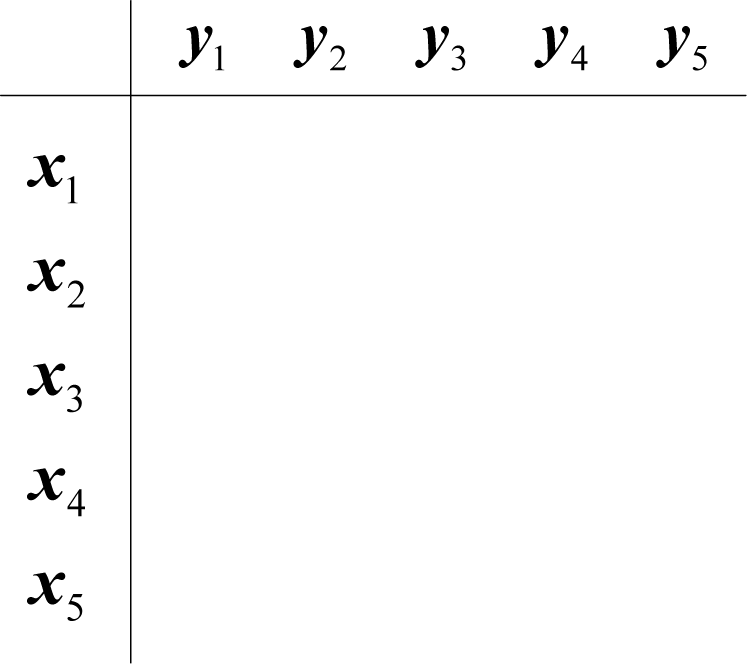}
    \includegraphics[width=0.325\hsize]{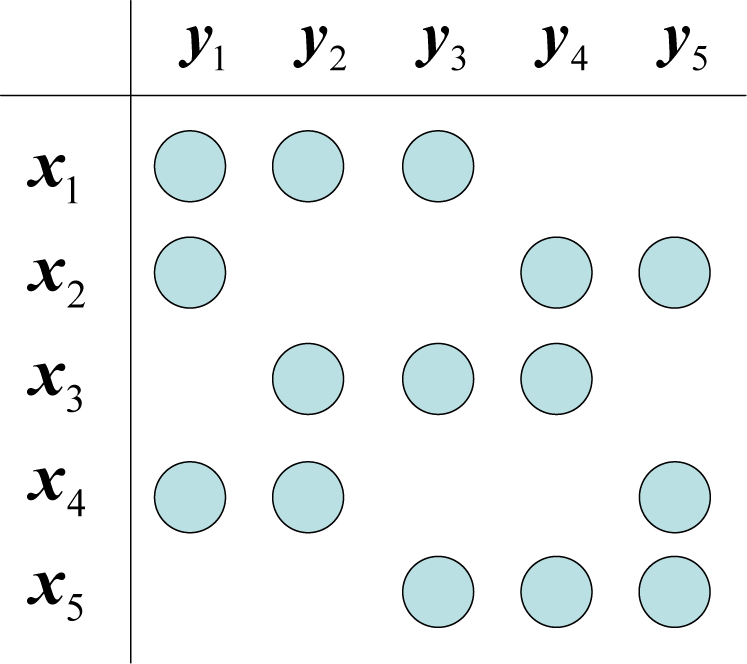}
    \includegraphics[width=0.325\hsize]{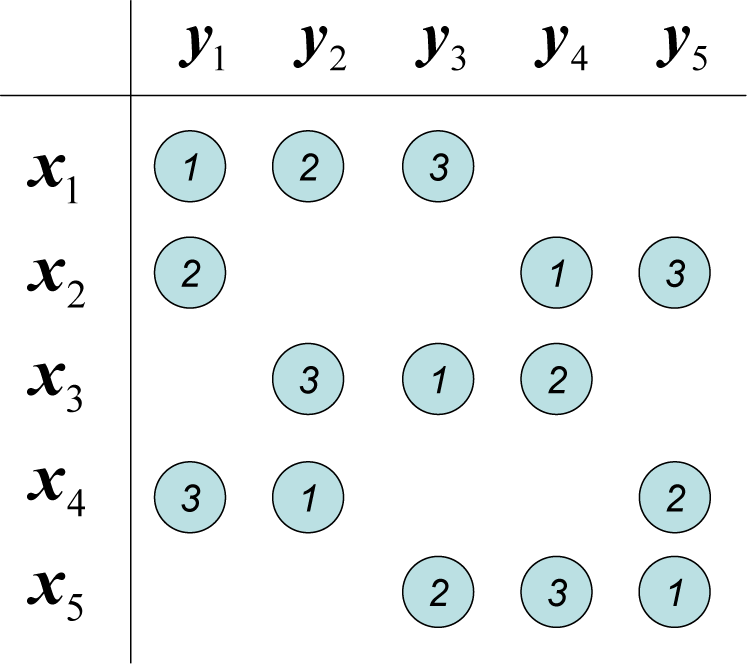}
    \caption{Example of coding scheme\quad
             (left)   Coding table\quad
             (middle) Positions where codewords will be provided\quad
             (right)  Provided codewords}
    \label{fig:CodingTable}
  \end{center}
\end{figure}

This section shows an explicit construction of universal codes for the complementary
delivery coding system defined by Definition \ref{def:code}. The coding scheme is
described as follows:

\noindent
[Encoding]
\begin{enumerate}
  \item Determine a set $\cS_n(R)$ of joint types as
        \begin{eqnarray*}
	  \cS_n(R)
	  &=& \{Q_{XY}\in\cP_n(\cX\times\cY):\\
	  & & \hspace{2mm}\max\{H(V|Q_X),H(W|Q_Y)\}\le R,\\
	  & & \hspace{2mm}Q_{XY}=Q_XV=Q_YW,\\
	  & & \hspace{2mm}V\in\cV_n(\cY|Q_X),W\in\cV_n(\cX|Q_Y)\},
	\end{eqnarray*}
	where $R>0$ is a given coding rate. We note that the joint type $Q_{XY}$
	specifies the types $Q_X$, $Q_Y$, and the conditional types $V$ and $W$.
  \item Create a table (henceforth we call this a {\it coding table}, see Figure
        \ref{fig:CodingTable} left) for each joint type $Q_{XY}\in\cS_n(R)$.
	Each row of the coding table corresponds to a sequence $\x\in T_{Q_X}^n$, and
	each column corresponds to a sequence $\y\in T_{Q_Y}^n$.
  \item Mark cells that correspond to sequence pairs $(\x,\y)\in T_{Q_{XY}}^n$
        (see Figure \ref{fig:CodingTable} middle). Codewords will be given only to
        sequence pairs that correspond to marked cells.
  \item Fill the marked cells with $\exp(nR)$ different symbols
        such that each symbol occurs at most once in each row and at most once in each
	column. An example of symbol filling is shown on the right in Figure
	\ref{fig:CodingTable} right.
  \item For a given pair of sequences $(\x,\y)\in\cX\times\cY$ with the joint type
        $Q_{XY}$, if $Q_{XY}\in\cS_n(R)$, the index assigned to
        the joint type $Q_{XY}$ of $(\x,\y)$ is the first part of the codeword, and the
	symbol filling the cell of $(\x,\y)$ in the coding table of $Q_{XY}$ is
	determined as the second part of the codeword. For the sequence pairs $(\x,\y)$
	whose joint type $Q_{XY}$ does not belong to $\cS_n(R)$, the corresponding
	codeword is determined arbitrarily and an encoding error is declared.
\end{enumerate}

\noindent
[Decoding: $\wh{\varphi}_{(1)}^n$] (Almost the same as for $\wh{\varphi}_{(2)}^n$)
\begin{enumerate}
  \item Find the coding table of the type $\wh{Q}_{XY}$ that corresponds to the first
        part of the received codeword. The decoder can find the coding table used in the
	encoding scheme if no encoding error occurs. In this case, $\wh{Q}_{XY}$ should
	be $Q_{XY}$.
  \item Find the cell filled with the second part of the received codeword from the
        column of the side information sequence $\y\in T_{Q_Y}^n$. The sequence
	$\wh{\x}\in T_{Q_X}^n$ that corresponds to the row of the cell found in this step
	is reproduced.
\end{enumerate}

\medskip
First, we show the existence of such coding tables. To this end, we introduce
the following two lemmas.

\begin{hodai} \label{lemma:table}
  For a given coding table of a joint type
  $Q_{XY}\in\cP_n(\cX\times\cY)$, the number of marked cells in every row of the coding
  table, $N_y(Q_{XY})$, is a constant value that is less than $\exp(nR)$, and the number
  of marked cells in every column of the coding table, $N_x(Q_{XY})$, is also a constant
  value of less than $\exp(nR)$, both of which depend solely on the joint type $Q_{XY}$. 
\end{hodai}

\begin{proof}
Note that the number of marked cells in each row equals the cardinality of the V-shell
$T_V^n(\x)$ for the sequence $\x\in T_{Q_X}^n$ that corresponds to the row. The
cardinality of V-shells $T_V^n(\x)$ is constant for a given joint type $Q_{XY}$ and any
sequences $\x\in T_{Q_X}^n$, this cardinality is bounded as follows:
\begin{eqnarray*}
  N_y(Q_{XY})
  & = & |T_V^n(\x)|\\
  &\le& \exp\{nH(V|Q_X)\}\quad(\because\mbox{Lemma \ref{lemma:sizeshell}})\\
  &\le& \exp(nR).
\end{eqnarray*}
In the same way, the number of marked cells in each column equals the cardinality of the
V-shell $T_W(\y)$ for the sequence $\y\in T_{Q_Y}^n$ that corresponds to the column, and
therefore it can be bounded as
\begin{eqnarray*}
  N_y(Q_{XY}) = |T_W^n(\y)| &\le& \exp(nR).
\end{eqnarray*}
This concludes the proof of Lemma \ref{lemma:table}.
\end{proof}

\begin{hodai} \label{lemma:graph}
  For given integers $m_x$, $m_y$, $n_x$ and $n_y$ that satisfy $m_x\ge n_x$ and
  $m_y\ge n_y$, there exists an $m_x\times m_y$ table filled with $\max(n_x,n_y)$
  different symbols such that
  \begin{itemize}
    \item at most $n_y$ cells are filled with a certain symbol for each row
          (blank cells are possible),
    \item at most $n_x$ cells are filled with a certain symbol for each column
          (blank cells are possible),
    \item each symbol occurs at most once in each row and at most once in each column.
  \end{itemize}
\end{hodai}

\begin{proof}
The table mentioned in this lemma is equivalent to a bipartite graph such that
\begin{itemize}
  \item each node in one set corresponds to a row in the table, and each node in the
        other set corresponds to a column in the table,
  \item each edge corresponds to a cell in the table, to which a certain symbol is
        assigned,
  \item $\max(n_x,n_y)$ different colors are given to edges, each of which corresponds to
        a symbol in the table,
  \item no two edges with the same color share a common node.
\end{itemize}
\begin{figure}[t]
  \begin{center}
    \includegraphics[width=0.85\hsize]{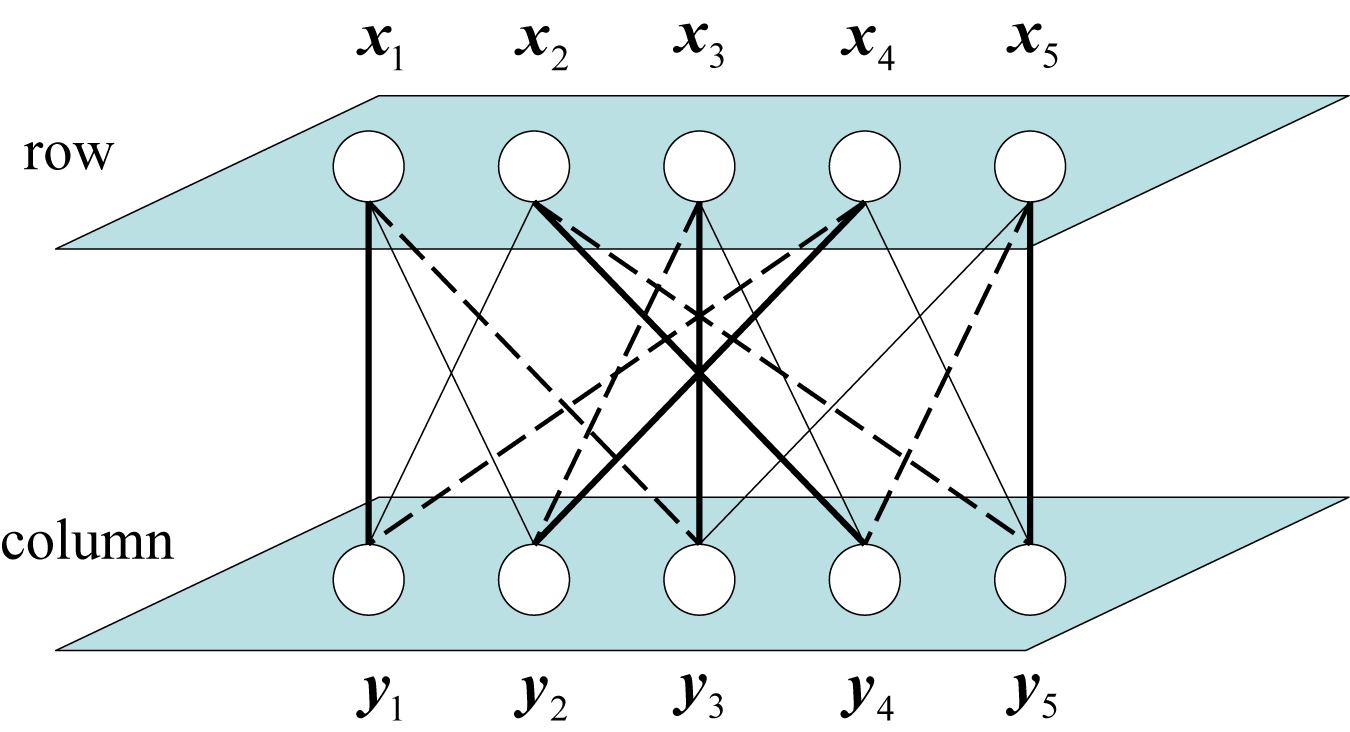}
    \caption{Example of a bipartite graph ($m_x=m_y=5$, $n_x=n_y=3$,
             equivalent to the table in Fig. \ref{fig:CodingTable} right)}
    \label{fig:BipartiteGraph}
  \end{center}
\end{figure}
Figure \ref{fig:BipartiteGraph} shows an example of such a graph. Here, let us introduce
the following lemma for bipartite graphs:
\begin{hodai} \label{lemma:konig}
  {\rm (K\dotdot{o}nig \cite{EdgeColoring,GraphTheory:biggs})}\\
  If a graph $G$ is bipartite, the minimum number of colors necessary for edge coloring
  of the graph $G$ equals the maximum degree of $G$.
\end{hodai}
\noindent
Lemma \ref{lemma:konig} ensures the existence of the above bipartite graph. This
concludes the proof of Lemma \ref{lemma:graph}.
\end{proof}

From Lemmas \ref{lemma:table} and \ref{lemma:graph}, we can easily show the existence of
coding tables by setting  $m_x=|T_{Q_X}^n|$, $m_y=|T_{Q_Y}^n|$, $n_x=|T_W^n(\y)|$ and
$n_y=|T_V^n(\x)|$ in Lemma \ref{lemma:graph}.

%%%%%%%%
\section{Coding theorems}
\label{sec:theorem}

We can obtain the following theorem for the universal FF-CD codes constructed in Section
\ref{sec:construct}. The proof is similar to that of the theorem of universal coding for
a single source.

\begin{teiri} \label{theorem:UniversalCode:direct}
  For a given real number $R>0$, there exists a sequence of universal FF-CD codes
  with parameters $\{(n,$ $M_n,$ $e_n^{(X)},$ $e_n^{(Y)})\}_{n=1}^{\infty}$ such
  that for any integer $n\ge 1$ and any source $(X,Y)$ with a generic distribution
  $P_{XY}\in\cM(\cX\times\cY)$
  \begin{eqnarray*}
    \frac 1n\log M_n &\le& R+\frac 1n|\cX\times\cY|\log(n+1),\\
    e_n^{(X)}+e_n^{(Y)}
    &\le& 2(n+1)^{|\cX\times\cY|}\\
    &   & \times\exp\left\{-n\hspace{-3mm}\min_{Q_{XY}\in\ov{\cS}_n(R)}
          \hspace{-3mm}D(Q_{XY}\| P_{XY})\right\},
  \end{eqnarray*}
  where $\ov{\cS}_n(R) = \cP_n(\cX\times\cY)-\cS_n(R)$.
\end{teiri}

\begin{proof}
Lemmas \ref{lemma:table} and \ref{lemma:graph} ensure the existence of a coding table for
every joint type $Q_{XY}\in\cP(\cX\times\cY)$. From the coding scheme, the size of the
codeword set is bounded as
\begin{eqnarray*}
  M_n &\le& |\cP_n(\cX\times\cY)|\exp(nR)\\
      &\le& (n+1)^{|\cX\times\cY|}\exp(nR),
            \quad(\because\mbox{Lemma \ref{lemma:typecount}})
\end{eqnarray*}
which implies the first inequality of Theorem
\ref{theorem:UniversalCode:direct}.
Next, we evaluate decoding error probabilities. Since every sequence pair
$(\x,\y)\in T_{Q_{XY}}^n$ that satisfies $Q_{XY}\in\cS_n(R)$ is reproduced correctly at
the decoder, the sum of error probabilities is bounded as
\begin{eqnarray*}
  \lefteqn{e_n^{(X)}+e_n^{(Y)}}\nonumber\\
  &\le& 2\Pr\left\{\exists Q_{XY}\in\ov{\cS}_n(R),~ (X^n,Y^n)\in T_{Q_{XY}}^n\right\}\\
  &\le& 2\sum_{Q_{XY}\in\ov{\cS}^n(R)}\exp\{-nD(Q_{XY}\| P_{XY})\}\\
  &   & \quad(\because\mbox{Lemma \ref{lemma:prob}})\\
  &\le& 2\sum_{Q_{XY}\in\ov{\cS}^n(R)}\hspace{-4mm}
        \exp\left\{-n\min_{Q_{XY}\in\ov{\cS}^n(R)}D(Q_{XY}\| P_{XY})\right\}\\
  &\le& 2(n+1)^{|\cX\times\cY|}\\
  &   & \times\exp\left\{-n\min_{Q_{XY}\in\ov{\cS}^n(R)}D(Q_{XY}\| P_{XY})\right\}\\
  &   & \quad(\because\mbox{Lemma \ref{lemma:typecount}})\nonumber
\end{eqnarray*}
This completes the proof of Theorem \ref{theorem:UniversalCode:direct}.
\end{proof}

\medskip
We can see that for any real value $R\ge R_f(X,Y)$ we have
\begin{eqnarray*}
  \min_{Q_{XY}\in\ov{\cS}_n(R)}D(Q_{XY}\|P_{XY}) &>& 0.
\end{eqnarray*}
This implies that any real value $R\ge R_f(X,Y)$ is
{\it a universal FF-CD achievable rate} of $(X,Y)$, namely, there exists a sequence of
universal FF-CD codes with parameters
$\{(n,M_n,e_n^{(X)},e_n^{(Y)})\}_{n=1}^{\infty}$ such that
\begin{eqnarray*}
  && \limsup_{n\to\infty}\frac 1n\log M_n \le R,\\
  && \lim_{n\to\infty}e_n^{(X)} = \lim_{n\to\infty}e_n^{(Y)} = 0.
\end{eqnarray*}

The following converse theorem indicates that the error exponent obtained in Theorem
\ref{theorem:UniversalCode:direct} is tight.

\begin{teiri} \label{theorem:UniversalCode:converse}
  Any sequence of FF-CD codes with parameters
  $\{(n,M_n,e_n^{(X)},e_n^{(Y)})\}_{n=1}^{\infty}$ for the source $(X,Y)$ must
  satisfy
  \begin{eqnarray*}
    \lefteqn{e_n^{(X)}+e_n^{(Y)} \ge \frac 12 (n+1)^{-|\cX\times\cY|}}\\
    & & \times\exp\left\{-n\min_{Q_{XY}\in\ov{\cS}_n(R+\epsilon_n)}D(Q_{XY}\| P_{XY})
        \right\}
  \end{eqnarray*}
  for any integer $n\ge 1$ and a given coding rate $R=1/n\log M_n>0$, where
  \begin{eqnarray*}
    \epsilon_n &{\displaystyle\mathop{=}^{\mbox{\rm def.}}}&
      \frac 1n \{|\cX\times\cY|\log(n+1)+1\}\\
    &\to& 0\quad(n\to\infty).
  \end{eqnarray*}
\end{teiri}

\begin{proof}
Note that the number of sequences to be decoded correctly for each decoder are at most
$\exp(nR)$. Here, let us consider a joint type $Q_{XY}\in\ov{\cS}_n(R+\epsilon_n)$. Lemma
\ref{lemma:sizeshell} and the definition of $\ov{\cS}_n(R+\epsilon_n)$ imply that for
$(\x,\y)\in T_{Q_{XY}}^n$ we have
\begin{eqnarray*}
  \lefteqn{\max\{|T_V^n(\x)|,|T_W^n(\y)|\}}\\
  &\ge& (n+1)^{-|\cX\times\cY|}\\
  &   & \times\max[\exp\{nH(V|Q_X)\},\exp\{nH(W|Q_Y)\}]\\
  &\ge& (n+1)^{-|\cX\times\cY|}\exp(n(R+\epsilon_n))\\
  & = & 2\exp(nR).
\end{eqnarray*}
Therefore, at least half of the sequences in the V-shell $T_V^n(\x)$ will not be decoded
correctly at the decoder $\varphi_{(2)}^n$, or at least half of sequences in the V-shell
$T_W^n(\y)$ will not be decoded correctly at the decoder $\varphi_{(1)}^n$. Thus, the
sum of error probabilities is bounded as
\begin{eqnarray*}
  \lefteqn{e_n^{(X)}+e_n^{(Y)}}\\
  &\ge& \frac 12\sum_{Q_{XY}\in\ov{\cS}_n(R+\epsilon_n)}\Pr\{(X^n,Y^n)\in T_{Q_{XY}}\}\\
  &\ge& \frac 12(n+1)^{-|\cX\times\cY|}\\
  &   & \times\sum_{Q_{XY}\in\ov{\cS}_n(R+\epsilon_n)}\exp\{-nD(Q_{XY}\| P_{XY})\}\\
  &   & \quad(\because\mbox{Lemma \ref{lemma:prob}})\\
  &\ge& \frac 12(n+1)^{-|\cX\times\cY|}\\
  &   & \times\exp\left\{-n\min_{Q_{XY}\in\ov{\cS}_n(R+\epsilon_n)}D(Q_{XY}\| P_{XY})
        \right\}
\end{eqnarray*}
This concludes the proof of Theorem \ref{theorem:UniversalCode:converse}.
\end{proof}

\medskip
The following corollary is directly derived from Theorems
\ref{theorem:UniversalCode:direct} and \ref{theorem:UniversalCode:converse}.

\begin{kei}
  For a given real number $R>0$, there exists a sequence of universal FF-CD codes with  
  parameters $\{(n,M_n,e_n^{(X)},e_n^{(Y)})\}_{n=1}^{\infty}$ such that for any
  source $(X,Y)$ 
  \begin{eqnarray*}
    \hspace{-5mm}&& \limsup_{n\to\infty}\frac 1n\log M_n \le R,\\
    \hspace{-5mm}&& \lim_{n\to\infty}-\frac 1n\log(e_n^{(X)}+e_n^{(Y)})
       = \hspace{-2mm}\min_{Q_{XY}\in\ov{\cS}(R)}\hspace{-2mm}D(Q_{XY}\| P_{XY}),
  \end{eqnarray*}
  where
  \begin{eqnarray*}
    \cS(R)
    &=& \{Q_{XY}\in\cM(\cX\times\cY):\\
    & & \hspace{2mm}\max\{H(V|Q_X),H(W|Q_Y)\}\le R,\\
    & & \hspace{2mm}Q_{XY}=Q_XV=Q_YW,\\
    & & \hspace{2mm}V\in\cM(\cY|Q_X),W\in\cM(\cX|Q_Y)\},
  \end{eqnarray*}
  and $\ov{\cS}(R) = \cM(\cX\times\cY)-\cS(R)$.
\end{kei}

In a similar manner, we can investigate the probability such that the original sequence
pair is correctly reproduced. The following theorem shows the lower bound of the
probability that can be attained by the proposed coding scheme.

\begin{teiri} \label{theorem:UniversalCode:lowrate:direct}
  For a given real number $R>0$, there exists a universal lossless f-FCD code
  $\{(\varphi_n,\wh{\varphi}_n^{(1)},\wh{\varphi}_n^{(2)})\}_{n=1}^{\infty}$
  such that for any integer $n\ge 1$ and any DMS $(X,Y)$
  \begin{eqnarray*}
    \frac 1n\log M_n &\le& R+\epsilon_n,\\
    1-(e_n^{(1)}+e_n^{(2)})
    &\ge& \exp\left\{-n\left(\epsilon_n+\min_{Q_{XY}\in\cT_n(R)}D(Q_{XY}\| P_{XY})
          \right)\right\}.
  \end{eqnarray*}
\end{teiri}

\begin{proof}
The first inequality is derived in the same way as the proof of Theorem
\ref{theorem:UniversalCode:direct}. Next, we evaluate the probability such that
the original sequence pair is correctly reproduced. Since every sequence pair
$(\x,\y)\in T_{Q_{XY}}^n$ that satisfies $Q_{XY}\in\cT_n(R)$ is reproduced correctly at
the decoder, the sum of the probabilities is bounded as
\begin{eqnarray}
  \lefteqn{1-(e_n^{(1)}+e_n^{(2)})}\nonumber\\
  &\ge& \Pr\left\{(X^n,Y^n)\in T_{Q_{XY}}^n: Q_{XY}\in\cT_n(R)\right\}
        \nonumber\\
  &\ge& \sum_{Q_{XY}\in\cT^n(R)}\hspace{-4mm}
        (n+1)^{-|\cX\times\cY|}\exp\{-nD(Q_{XY}\| P_{XY})\}
        \label{eq:proof:UniversalCode:lowrate:direct:1}\\
  &\ge& (n+1)^{-|\cX\times\cY|}
        \exp\left\{-n\min_{Q_{XY}\in\cT^n(R)}D(Q_{XY}\| P_{XY})\right\}\nonumber\\
  & = & \exp\left\{-n\left(\epsilon_n+\min_{Q_{XY}\in\cT_n(R)}D(Q_{XY}\| P_{XY})
        \right)\right\},\nonumber
\end{eqnarray}
where Eq. (\ref{eq:proof:UniversalCode:lowrate:direct:1}) comes from Lemma
\ref{lemma:prob}.
This completes the proof of Theorem \ref{theorem:UniversalCode:lowrate:direct}.
\end{proof}

The following converse theorem indicates that the error exponent obtained in Theorem
\ref{theorem:UniversalCode:lowrate:direct} might not be tight.

\begin{teiri} \label{theorem:UniversalCode:lowrate:converse}
  Any lossless f-FCD code
  $\{(\varphi_n,\wh{\varphi}_n^{(1)},\wh{\varphi}_n^{(2)})\}_{n=1}^{\infty}$
  for the DMS $(X,Y)$ must satisfy
  \begin{eqnarray*}
    \lefteqn{1-(e_n^{(1)}+e_n^{(2)})}\\
    &\le& \exp\Bigl\{-n\Bigl(-\epsilon_n+\min_{Q_{XY}\in\cP_n(\cX\times\cY)}\\
    &   & \left|\max(H(V|Q_X),H(W|Q_Y))-\left(R+\epsilon_n\right)\right|^{+}
	  +D(Q_{XY}\| P_{XY})\Bigr)\Bigr\}
  \end{eqnarray*}
  for any integer $n\ge 1$ and a given coding rate $R=1/n\log M_n>0$.
\end{teiri}

\begin{proof}
Note that the number of sequences to be decoded correctly for each decoder are at most
$\exp(nR)$. Here, let us consider a joint type $Q_{XY}\in\cP_n(\cX\times\cY)$ such that
$Q_{XY}=Q_XV=Q_YW$, $V\in\cM(\cY|Q_X)$ and $W\in\cM(\cX|Q_Y)\}$. The ratio $r_c(Q_{XY})$
of sequences in the sequence set $T_{Q_{XY}}$ such that the sequences are correctly
reproduced is at most
\begin{eqnarray}
  \lefteqn{r_c(Q_{XY})}\nonumber\\
  &\le& \min\left\{\min\left(\frac{\exp(nR)}{|T_V^n(\x)|},\frac{\exp(nR)}{|T_W^n(\y)|}
        \right),1\right\}\nonumber\\
  &\le& \min\Bigl[\exp(nR)\cdot(n+1)^{|\cX\times\cY|}
	\exp\{-n\max(H(V|Q_X),H(W|Q_Y))\},1\Bigr]\\
  &   & \label{eq:proof:lowrate:converse:1}\\
  & = & \min\Bigl[\exp\left\{n\left(R+\gamma_n\right)\right\}
	\exp\{-n\max(H(V|Q_X),H(W|Q_Y))\},1\Bigr] \nonumber\\
  & = & \exp\left\{-n\left|\max\{H(V|Q_X),H(W|Q_Y)\}-\left(R+\gamma_n\right)\right|^{+}
	\right\} \nonumber
\end{eqnarray}
where Eq. (\ref{eq:proof:lowrate:converse:1}) comes from Lemma
\ref{lemma:sizeshell}.
Therefore, the probability $P_c(Q_{XY})$ such that the original sequence pair with type
$Q_{XY}$ is correctly reproduced is bounded as
\begin{eqnarray*}
  \lefteqn{P_c(Q_{XY})}\\
  &\le& r_c(Q_{XY})\Pr\{(X^n,Y^n)\in T_{Q_{XY}}^n\}\\
  &\le& \exp\bigl\{-n\bigl|\max(H(V|Q_X),H(W|Q_Y))
        -\left(R+\gamma_n\right)\bigr|^{+}+D(Q_{XY}\| P_{XY})\bigr\}.
\end{eqnarray*}
Thus, the sum of the probabilities is obtained as
\begin{eqnarray*}
  \lefteqn{1-(e_n^{(1)}+e_n^{(2)})}\\
  &\le& \sum_{Q_{XY}\in\cP_n(\cX\times\cY)}P_c(Q_{XY})\\
  &\le& \sum_{Q_{XY}\in\cP_n(\cX\times\cY)}\hspace{-6mm}\exp\bigl[-n
        \bigl|\max\{H(V|Q_X),H(W|Q_Y)\}
	-\left(R+\gamma_n\right)\bigr|^{+}+D(Q_{XY}\| P_{XY})\bigr]\\
  &\le& \exp\bigl[-n\min_{Q_{XY}\in\cP_n(\cX\times\cY)}
        \bigl(\bigl|\max\{H(V|Q_X),H(W|Q_Y)\}\\
  &   & \hspace{10mm}-\left(R+\gamma_n\right)\bigr|^{+}+D(Q_{XY}\| P_{XY})\bigr)\bigr].
\end{eqnarray*}
This completes the proof of Theorem
\ref{theorem:UniversalCode:lowrate:converse}.
\end{proof}

\medskip
We can see that for any real value $R\ge R_f(X,Y)$ we have
\begin{eqnarray*}
  \lefteqn{\min_{Q_{XY}\in\cM(\cX\times\cY)}\bigl[\bigl|\max\{H(V|Q_X),H(W|Q_Y)\}
           -R\bigr|^{+}+D(Q_{XY}\| P_{XY})\bigr)\bigr]}\\
  & = & \bigl|\max(H(P_{Y|X}|P_X),H(P_{X|Y}|P_Y))-R\bigr|^{+}\hspace{25mm}\\
  & = & 0.
\end{eqnarray*}
On the other hand, for any real value $R<R_f(X,Y)$ we have
\begin{eqnarray*}
  \lefteqn{\min_{Q_{XY}\in\cT(R)}D(Q_{XY}\| P_{XY})}\\
  &\ge& \min_{Q_{XY}\in\cM(\cX\times\cY)}\bigl(\bigl|\max(H(V|Q_X),H(W|Q_Y))
        -R\bigr|^{+}+D(Q_{XY}\| P_{XY})\bigr)\bigr\}\\
  &\ge& 0.
\end{eqnarray*}
This implies that the error exponent obtained in Theorem
\ref{theorem:UniversalCode:lowrate:direct} might not be tight.

%%%%%%%%
\section{Variable-length coding}
\label{sec:variable}

This section discusses variable-length coding for the complementary delivery coding
system, and shows an explicit construction of universal variable-length codes. The coding
scheme is similar to that of fixed-length codes, and also utilizes the coding tables
defined in Section \ref{sec:construct}.

%%%%
\subsection{Formulation}
\label{sec:variable:formulate}

\begin{teigi}  \label{def:code:variable}
  {\rm(Fixed-to-variable complementary delivery (FV-CD) code)}\\
  A set $(\varphi^n,\wh{\varphi}_{(1)}^n,\wh{\varphi}_{(2)}^n)$ of an encoder and two
  decoders is an FV-CD code for the source $(X,Y)$ if and only if
  \begin{eqnarray*}
    \varphi^n            &:& \cX^n\times\cY^n\rightarrow\cB^*\\
    \wh{\varphi}_{(1)}^n &:& \varphi^n(\cX^n,\cY^n)\times\cY^n
                             \rightarrow\cX^n,\\
    \wh{\varphi}_{(2)}^n &:& \varphi^n(\cX^n,\cY^n)\times\cX^n
                             \rightarrow\cY^n,\\
    e_n^{(X)} &=& \Pr\left\{X^n\neq\wh{X}^n\right\}=0,\\
    e_n^{(Y)} &=& \Pr\left\{Y^n\neq\wh{Y}^n\right\}=0,
  \end{eqnarray*}
  where
  \begin{eqnarray*}
    \wh{X}^n &{\displaystyle\mathop{=}^{\mbox{\rm def.}}}&
      \wh{\varphi}_{(1)}^n(\varphi^n(X^n,Y^n),Y^n),\\
    \wh{Y}^n &{\displaystyle\mathop{=}^{\mbox{\rm def.}}}&
      \wh{\varphi}_{(2)}^n(\varphi^n(X^n,Y^n),X^n),
  \end{eqnarray*}
  and the image of $\varphi^n$ is a prefix set.
\end{teigi}

\begin{teigi}  \label{def:rate:variable}
  {\rm(FV-CD-achievable rate)}\\
  $R$ is an FV-CD-achievable rate of the source $(X,Y)$ if and only if there exists a
  sequence of FV-CD codes $\{(\varphi^n,\wh{\varphi}_{(1)}^n,\wh{\varphi}_{(2)}^n)\}
  _{n=1}^{\infty}$ for the source $(X,Y)$ such that
  \begin{eqnarray*}
    \limsup_{n\to\infty}\frac 1n E\left[l(\varphi^n(X^n,Y^n))\right] &\le& R,
  \end{eqnarray*}
  where $l(\cdot): \cB^*\to\cR$ is a length function.
\end{teigi}

\begin{teigi}  \label{def:min_rate:variable}
  {\rm(Inf FV-CD-achievable rate)}
  \begin{eqnarray*}
    \lefteqn{R_v(X,Y) = \inf\{R:}\\
    & & R \mbox{ is an FV-CD-achievable rate of }(X,Y)\}.
  \end{eqnarray*}
\end{teigi}

%%%%
\subsection{Code construction}
\label{sec:variable:construct}

We can construct universal FV-CD codes in a similar manner to universal FF-CD codes. Note
that the coding rate depends on the type of sequence pair to be encoded, whereas the
coding rate is fixed beforehand for fixed-length coding. The coding scheme is described
as follows:

\noindent
[Encoding]
\begin{enumerate}
  \item Create a coding table for each joint type $Q_{XY}\in\cP_n(\cX\times\cY)$ in the
        same way as Step 2 of Section \ref{sec:construct}.
  \item Mark cells that correspond to sequence pairs $(\x,\y) \in T_{Q_{XY}}$.
  \item Fill the marked cells on the coding table with different
        $\max\{|T_V^n(\x)|,|T_W^n(\y)|\}$ symbols such that each symbol occurs at most
        once in each row and at most once in each column, where $\x\in T_{Q_X}^n$,
	$\y\in T_{Q_Y}^n$.
  \item For a given pair of sequences $(\x,\y)\in\cX^n\times\cY^n$, the number (index)
        assigned to the joint type
        $Q_{XY}$ of $(\x,\y)$ is the first part of the codeword, and the symbol filling
	the cell of $(\x,\y)$ in the coding table of $Q_{XY}$ is determined as the second
	part of the codeword.
\end{enumerate}

\noindent
[Decoding]\\
Decoding can be accomplished in almost the same way as the fixed-length coding. Note that
the decoder can always find the coding table used in the encoding scheme.

%%%%
\subsection{Coding theorems}
\label{sec:variable:theorem}

We begin by showing a theorem for (non-universal) variable-length coding, which
indicates that the inf coding rate of variable-length coding is the same as that of
fixed-length coding.

\begin{teiri} \label{theorem:complementary:variable}
  {\rm (Coding theorem of FV-CD code)}
  \begin{eqnarray*}
    R_v(X,Y) &=& R_f(X,Y) = \max\{H(X|Y),H(Y|X)\}.
  \end{eqnarray*}
\end{teiri}

\begin{proof}
See Appendix.
\end{proof}

The following direct theorem for universal coding indicates that the coding scheme
presented in the previous subsection can achieve the inf achievable rate
clarified in Theorem \ref{theorem:complementary:variable}.

\begin{teiri} \label{theorem:UniversalCode:variable:direct}
  There exists a sequence of universal FV-CD codes
  $\{(\varphi^n,\wh{\varphi}_{(1)}^n,\wh{\varphi}_{(2)}^n)\}_{n=1}^{\infty}$ such that
  for any integer $n\ge 1$ and any source $(X,Y)$, the overflow probability, namely the
  probability that the length of a codeword exceeds a given real number $R>0$, is
  bounded as
  \begin{eqnarray*}
    \ov{\rho}_n(R) &{\displaystyle\mathop{=}^{\mbox{\rm def.}}}&
    \Pr\left\{l(\varphi^n(X^n,Y^n)) > n(R+\epsilon_n)\right\}\\
    &\le& (n+1)^{|\cX\times\cY|}\\
    &   & \times\exp\left\{-n\min_{Q_{XY}\in\ov{\cS}_n(R)}
          \hspace{-3mm}D(Q_{XY}\| P_{XY})\right\},
  \end{eqnarray*}
  where $\epsilon_n$ is defined in Theorem \ref{theorem:UniversalCode:converse}. This
  implies that there exists a sequence of universal FV-CD codes
  $\{(\varphi^n,\wh{\varphi}_{(1)}^n,\wh{\varphi}_{(2)}^n)\}_{n=1}^{\infty}$ that
  satisfies
  \begin{eqnarray}
    \limsup_{n\to\infty}\frac 1n l(\varphi^n(X^n,Y^n)) &\le& R_v(X,Y)\quad\mbox{a.s.}
    \label{eq:theorem:variable:direct:1}
  \end{eqnarray}
\end{teiri}

\begin{proof}
The overflow probability can be obtained in the same way as an upperbound of the error
probability of FF-CD codes, which has been shown in the proof of Theorem
\ref{theorem:UniversalCode:direct}. Thus, from Theorem
\ref{theorem:complementary:variable} we have
\begin{eqnarray*}
  \sum_{n=1}^{\infty}\Pr\left\{\frac 1n l(\varphi^n(X^n,Y^n))>R_v(X,Y)+\delta\right\}
  & < & \infty
\end{eqnarray*}
for a given $\delta>0$. From Borel-Cantelli's lemma \cite[Lemma 4.6.3]{GrayBook},
we immediately obtain Eq.(\ref{eq:theorem:variable:direct:1}).
This completes the proof of Theorem \ref{theorem:UniversalCode:variable:direct}.
\end{proof}

\medskip\noindent
The following converse theorem for variable-length codingindicates
that the exponent of the overflow probability obtained in Theorem
\ref{theorem:UniversalCode:variable:direct} is tight. This can be easily obtained in
almost the same way as Theorem \ref{theorem:UniversalCode:converse}.

\begin{teiri} \label{theorem:UniversalCode:variable:converse}
  Any sequence of FV-CD codes
  $\{(\varphi^n,\wh{\varphi}_{(1)}^n$, $\wh{\varphi}_{(2)}^n)\}_{n=1}^{\infty}$
  for the source $(X,Y)$ must satisfy
  \begin{eqnarray*}
    \lefteqn{\ov{\rho}_n(R) \ge (n+1)^{-|\cX\times\cY|}}\\
    & & \times\exp\left\{-n\min_{Q_{XY}\in\ov{\cS}_n(R+\epsilon_n)}D(Q_{XY}\| P_{XY})
        \right\}.
  \end{eqnarray*}
  for a given real number $R>0$ and any integer $n\ge 1$, where $\epsilon_n$ is defined
  in Theorem \ref{theorem:UniversalCode:converse}.
\end{teiri}

\medskip
The following corollary is directly derived from Theorems
\ref{theorem:UniversalCode:variable:direct} and
\ref{theorem:UniversalCode:variable:converse}.
\begin{kei}
  There exists a sequence of universal FV-CD codes
  $\{(\varphi^n,\wh{\varphi}_{(1)}^n,\wh{\varphi}_{(2)}^n)\}_{n=1}^{\infty}$ such that
  for any source $(X,Y)$
  \begin{eqnarray*}
    && \limsup_{n\to\infty}\frac 1n l(\varphi^n(X^n,Y^n))
    \le R_v(X,Y)\quad\mbox{a.s.}\\
    && \lim_{n\to\infty}-\frac 1n\log\ov{\rho}_n(R)
    = \min_{Q_{XY}\in\ov{\cS}(R)}\hspace{-3mm}D(Q_{XY}\| P_{XY})
  \end{eqnarray*}
\end{kei}

Next, we investigate the underflow probability, namely the probability that the length of
a codeword falls below a given real number $R>0$. For this purpose, we present the
following two theorems. The proofs are almost the same as those of Theorems
\ref{theorem:UniversalCode:lowrate:direct} and
\ref{theorem:UniversalCode:lowrate:converse}.

\begin{teiri} \label{theorem:UniversalCode:variable:under:direct}
  There exists a universal lossless v-FCD code
  $\{(\varphi_n,\wh{\varphi}_n^{(1)},\wh{\varphi}_n^{(2)})\}_{n=1}^{\infty}$ such that
  for any integer $n\ge 1$ and any DMS $(X,Y)$, the underflow probability
  $\ul{\rho}_n(R)$ is bounded as
  \begin{eqnarray*}
    \ul{\rho}_n(R) &\eqdef& \Pr\left\{l(\varphi_n(X^n,Y^n)) < nR\right\}\\
    &\le& \exp\left\{-n\left(\epsilon_n+\min_{Q_{XY}\in\cT_n(R-\epsilon_n(2))}
          \hspace{-3mm}D(Q_{XY}\| P_{XY})\right)\right\}.
  \end{eqnarray*}
  This implies that there exists a universal lossless v-FCD code
  $\{(\varphi_n,\wh{\varphi}_n^{(1)},\wh{\varphi}_n^{(2)})\}_{n=1}^{\infty}$ that
  satisfies
  \begin{eqnarray}
    \liminf_{n\to\infty}\frac 1n l(\varphi_n(X^n,Y^n)) &\ge& R_v(X,Y)\quad\mbox{a.s.}
    \label{eq:theorem:variable:under:direct:1}
  \end{eqnarray}
\end{teiri}

\begin{teiri} \label{theorem:UniversalCode:variable:under:converse}
  Any lossless v-FCD code
  $\{(\varphi_n,\wh{\varphi}_n^{(1)}$, $\wh{\varphi}_n^{(2)})\}_{n=1}^{\infty}$
  for the DMS $(X,Y)$ must satisfy
  \begin{eqnarray*}
    \lefteqn{\ul{\rho}_n(R)}\\
    &\le& \exp\Bigl\{-n\Bigl(-\epsilon_n(1)+\min_{Q_{XY}\in\cP_n(\cX\times\cY)}\\
    &   & \left|\max(H(V|Q_X),H(W|Q_Y))-\left(R+\epsilon_n\right)\right|^{+}
          +D(Q_{XY}\| P_{XY})\Bigr)\Bigr\}
  \end{eqnarray*}
  for a given real number $R>0$ and any integer $n\ge 1$.
\end{teiri}

%%%%%%%%
\section{Concluding remarks}
\label{sec:conclude}

We invesigated a universal coding problem for the complementary delivery coding system.
First, we presented an explicit construction of universal fixed-length codes, which was
based on a graph-theoretical technique. We clarified that the error exponent achieved by
the proposed coding scheme is asymptotically optimal. Next, we applied the coding
scheme to construction of universal variale-length codes. We clarified that there exists
a universal code such that the codeword length converges to the minimum achievable rate
almost surely, and that the exponent of the overflow probability achieved by the proposed
coding scheme is optimal. This paper dealt with only the lossless configuration, and
therefore constructing universal lossy codes for the complementary delivery coding system
still remains as an open problem.

%%%%%%%%
\section*{Acknowledgements}
The authors would like to thank Dr. Jun Muramatsu of NTT Communication Science
Laboratories, and Prof. Ryutaroh Matsumoto of Tokyo Institute of Technology for their
valuable discussions and helpful comments, which led to improvements in this work. The
authors also thank Drs. Yoshinobu Tonomura, Hiromi Nakaiwa, Tatsuto Takeuchi, Shoji
Makino and Junji Yamato of NTT Communication Science Laboratories for their help.

%%%%%%%%
\appendix
%%%%
\section{Proof of Theorem \ref{theorem:complementary:variable}}
\subsection{Direct part}
\begin{proof}
We can apply a sequence of {\it achievable} FF-CD codes (fixed-length codes). The encoder
$\varphi^n$ assigns the same codeword as that of the fixed-length code to a sequence pair
$(\x,\y)\in\cX^n\times\cY^n$ that is correctly reproduced by the fixed-length
code. Otherwise, the encoder sends the sequence pair itself as a codeword.

The above FV-CD code can always reproduce the original sequence pair at the decoders, and
it attains the desired coding rate.
\end{proof}

\subsection{Converse part}
\begin{proof}
We can prove the converse part in a similar manner to that for fixed-length
coding. Let a sequence $\{(\varphi^n,\wh{\varphi}_{(1)}^n,\wh{\varphi}_{(2)}^n)\}_{n=1}
^{\infty}$ of FV-CD codes be given that satisfies the conditions of Definitions
\ref{def:code:variable} and \ref{def:rate:variable}. From Definition
\ref{def:rate:variable}, for any $\delta>0$ there exists an integer $n_1=n_1(\delta)$ and
then for all $n\ge n_1(\delta)$, we can obtain
\begin{eqnarray}
  \frac 1n E[l(\varphi^n(X^n,Y^n))] &\le& R+\delta.
  \label{eq:proof:fv:converse:1}
\end{eqnarray}
Here, let us define $A_n=\varphi^n(X^n,Y^n)$. Since the decoder $\wh{\varphi}_{(1)}^n$
can always reproduce the original sequence $X^n$ from the received codeword $A_n$ and
side information $Y^n$, we can see that
\begin{eqnarray}
  H(X^n|A_nY^n) &=& 0.  \label{eq:proof:fv:converse:2}
\end{eqnarray}
Similarly, we can obtain
\begin{eqnarray*}
  H(Y^n|A_nX^n) &=& 0.
\end{eqnarray*}
Substituting $A_n$ into Eq.(\ref{eq:proof:fv:converse:1}), we have
\begin{eqnarray*}
  n(R+\delta)
  &\ge& E[l(A_n)]\\
  &\ge& H(A_n)\quad(\because A_n\mbox{ is a prefix set})\\
  &\ge& H(A_n|Y^n)\\
  &\ge& I(X^n;A_n|Y^n)\\
  & = & H(X^n|Y^n).\quad(\because\mbox{Eq.(\ref{eq:proof:fv:converse:2})})
\end{eqnarray*}
Since we can select an arbitrarily small $\delta>0$ for a sufficient large $n$, we can
obtain
\begin{eqnarray*}
  R &\ge& \frac 1n H(X^n|Y^n) = H(X|Y).
\end{eqnarray*}
In the same way, we also obtain
\begin{eqnarray*}
  R &\ge& H(Y|X).
\end{eqnarray*}
\end{proof}

%%%%%%%%
\bibliographystyle{bib/ieicetr}
\bibliography{bib/IEEEabrv,bib/defs,bib/it}

\end{document}